\documentclass[prl,twocolumn,superscriptaddress,amsmath,amssymb]{revtex4-1}
\newcommand{\fatr}{\mathbf{r}}

\usepackage{mhchem}
\usepackage{threeparttable}
\usepackage{array}
\usepackage{graphicx}
\usepackage{color}
\usepackage{dcolumn}
\usepackage{bm}
\usepackage[final]{pdfpages}
\usepackage{float}
\usepackage{footnote}

\def\beq{\begin{equation}}
\def\eeq{\end{equation}}
\def\bea{\begin{eqnarray}}
\def\eea{\end{eqnarray}}

\def\fatr{{\bf r}}

\begin{document}
\title{Quantum Mechanical Treatment of Variable Molecular Composition: 
From ``Alchemical'' Changes of State Functions to Rational Compound Design} 
\author{K. Y. Samuel Chang}
\affiliation{Institute of Physical Chemistry, Department of Chemistry, University of Basel, 4056 Basel, Switzerland}
\author{O. Anatole von Lilienfeld}
\email{anatole.vonlilienfeld@unibas.ch}
\affiliation{Institute of Physical Chemistry, Department of Chemistry, University of Basel, 4056 Basel, Switzerland}
\affiliation{Argonne Leadership Computing Facility, Argonne National Laboratory, Argonne, IL 0439, USA}

\date{\today}

\begin{abstract}
\noindent 
\noindent
``Alchemical'' interpolation paths, i.e.~coupling systems along fictitious paths that
without realistic correspondence, are frequently used within materials and molecular
modeling and simulation protocols for the estimation of relative changes in
state functions such as free energies.
We discuss alchemical changes in the context of quantum chemistry,
and present illustrative numerical results for the changes of HOMO eigenvalues of the He atom
due to a linear alchemical teleportation---the simultaneous annihilation and creation of nuclear charges at different locations.
To demonstrate the predictive power of alchemical first order derivatives (Hellmann-Feynman)
the covalent bond potential of hydrogen fluoride and hydrogen chloride is investigated, as well as
the van-der-Waals binding in the water-water and water-hydrogen fluoride dimer, respectively.
Based on converged electron densities for one configuration, the versatility of
alchemical derivatives is exemplified for the screening of entire binding potentials with reasonable accuracy.
Finally, we discuss constraints for the identification of non-linear coupling potentials
for which the energy's Hellmann-Feynman derivative will yield accurate predictions.
\end{abstract}

\maketitle

\section{Introduction}
\begin{figure}
\centering
\includegraphics[scale=0.3, angle=0]{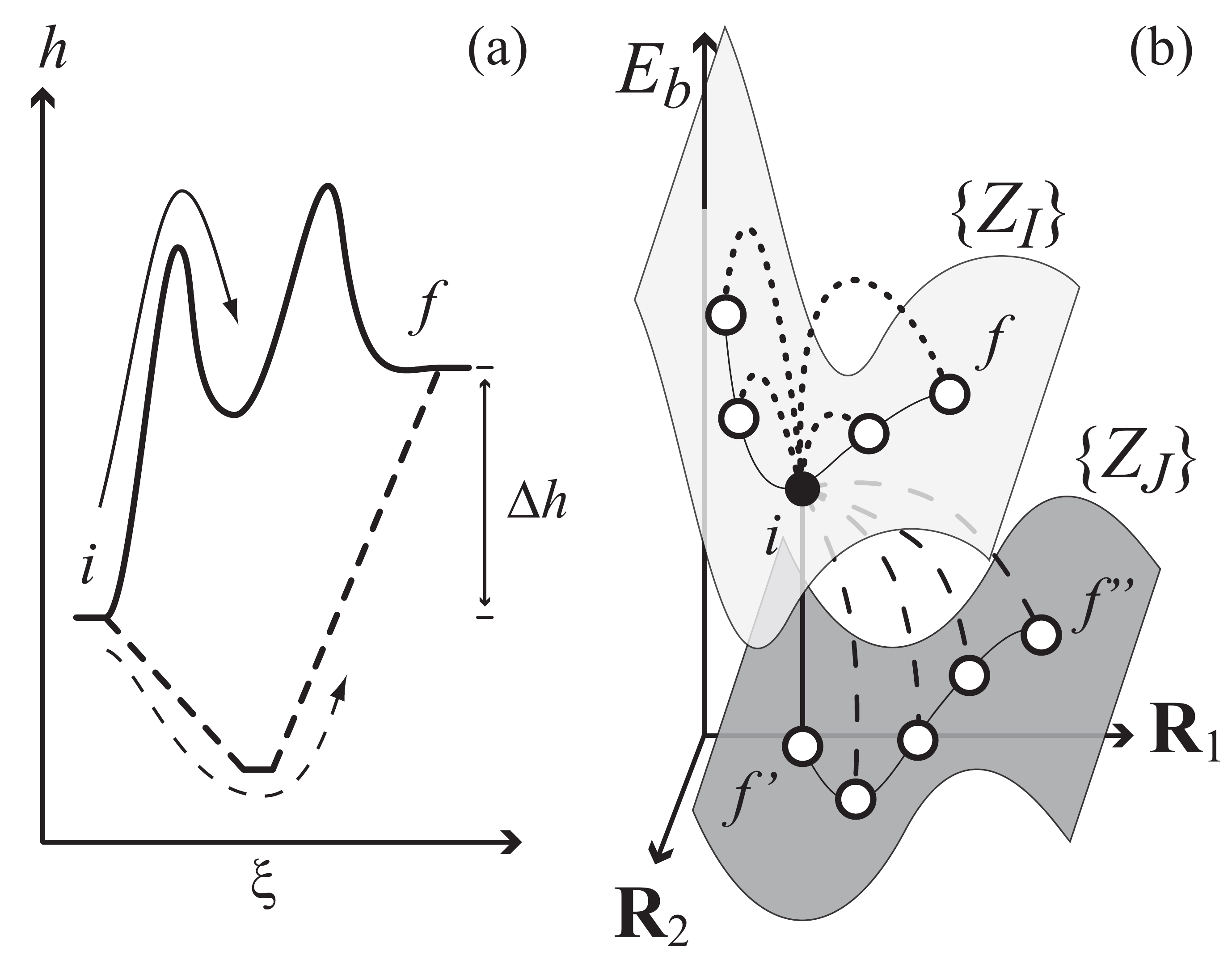}
\caption{Alchemical cartoons.
(a) The same enthalpy change, $\Delta h$, is obtained for a realistic (full) or 
an alchemical (dashed) coupling between initial ($i$) and final ($f$) states 
as a function of reaction progress $\xi$.
(b) Alchemical paths connecting compounds on two two-dimensional binding potential energy 
surfaces corresponding to two different stoichiometries, $\{Z_I\}$ (white) and $\{Z_J\}$ (gray), respectively.
Having calculated $E_b$ for some initial system $i$ (filled circle),
alchemical paths couple to energies (open circle) of different geometries $f$ with same stoichiometry (dotted), 
different stoichiometries with same geometry $f'$ (full), 
or different geometries and stoichiometries $f''$ (dashed). 
}
\label{fig:enthalpy}
\label{fig:alchemicalPath}
\end{figure}

Ever since the introduction of Hess' law and Carnot's cycle, 
chemists have known that some properties, called state functions, always change by the same amount
when a system is moved reversibly from one state to another---regardless of how the change has been implemented.
The freedom to choose {\em any} paths, even 
paths without any realistic correspondence except for the endpoints, 
is exploited within many applications.
We generally refer to ``alchemical'' paths as paths that cannot be followed and verified through experimental observations. 
For example, Fig.~\ref{fig:enthalpy}(a) illustrates how, according to Hess' law, 
the change of enthalpy of reaction can be calculated either by following
the (realistic) reaction path, or, just as well, by following a more convenient yet non-realistic (alchemical)
reaction path to product via dissembled elemental states as intermediates.
Depending on the choice of state function, external conditions, system, and process,
realistic reaction paths can be significantly more challenging 
because they can involve many intermediate and transition states which are difficult to identify and characterize. 
Even worse, they might even be experimentally impossible to probe, 
as it is the case for the chemistry of the earth's core, 
some other planet's bio-sphere, 
for distant historical or future events, 
or for very slow or very fast processes.

Within the atomistic theories of quantum and statistical mechanics, {\em any} path connecting the Hamiltonian of 
some initial molecule or material system, $H_i$, to some final system $H_f$, can be defined 
in a coupling order parameter $\lambda$ as long as as the end-points are met,~\cite{Gunsteren_jcamd_1987,Straatsma_arpc_1992,Kollman_chemrev_1993}
i.e.~
\begin{equation}\label{eq:H_lambda}
	H(\lambda) = 
	\left\lbrace
	\begin{array}{lc}
		H_i, & \lambda=0,\\[4pt]
		H_{\lambda}, & 0<\lambda<1,\\[4pt]
		H_f, & \lambda=1,
	\end{array}
	\right.
\end{equation}
where $0\leq\lambda \leq 1$.
$H_{\lambda}$ in Eq.~(\ref{eq:H_lambda}) denotes some intermediate state at $\lambda$, not necessarily differentiable. 
At boundaries of first order phase transitions, for example, the entropy (state function) is not continuous in temperature ($\lambda$). 
Often, $H(\lambda)$ is (arbitrarily) chosen to be linear in $\lambda$, i.e.  $H(\lambda) = H_i + \lambda (H_f - H_i)$. 
As alluded to above, $H_{\lambda}$ does not have to be realistic for all values of $\lambda$.
Thermodynamics textbook examples of such changes include the calculation of the errors made when relying on the ideal gas equation.
Introduced as ``computational alchemy''\cite{Straatsma_arpc_1992,Kollman_chemrev_1993} in the realm of computational
chemistry, this concept has successfully been used for the interpolation of forces and energies for molecular dynamics (MD) and Monte Carlo (MC) simulations. 
Also for the purpose of quantum mechanical observables, we can denote any such unrealistic path as 
``alchemical''.\cite{Marzari_prl_1994,Anatole_prl_2005, Anatole_jcp_2006} 
We note however that Eq.~(\ref{eq:H_lambda}) is also known as ``mutation path'' or ``adiabatic connection''.\cite{Jorgensen_jcp_1985, Perdew_prb_1980, Harris_pra_1984}

An even more intriguing possibility for exploiting the freedom of alchemical changes relates to the challenge of rational compound design (RCD).
RCD attempts to circumvent (or at least reduce) the combinatorially scaling challenge of having to virtually enumerate and 
screen larger subsections of chemical or materials compound space using computationally demanding simulation methods. 
It has already been shown to yield promising results for the virtual atomistic control of material, nanoparticle, and even molecular structures.\cite{exp_nature_2005, exp_acie_2009} 
Because of the vastness of chemical compound space (CCS), identification of novel compounds that meet desired property requirements still remains a challenge.\cite{Anatole_prl_2005, Anatole_ijqc_2013} 
Once an alchemical interpolating path, $H(\lambda)$, is defined, property derivatives with respect to $\lambda$ 
can be evaluated \cite{Anatole_jcp_2009} (see Sec.~\ref{sec:currentWork}). 
Similar to an iterative gradient descent-like algorithm, one can thus navigate gigantic combinatorial compound libraries at dramatically 
reduced computational costs by visiting the most promising compounds one after the other while avoiding the least promising candidates.\cite{Anatole_jcp_2010, Minimahopping} 

The concept of connecting different systems via Eq.~(\ref{eq:H_lambda}) has been in frequent use in various research fields, including
computational engineering, physics, biophysics, and chemistry. 
Here, we first briefly summarize the most common application of Eq.~(\ref{eq:H_lambda}) 
to calculate free energy changes, or alloy formation energies in Sec.~\ref{sec:statistical}.
In Sec.~\ref{sec:QMAlchemy} we review the quantum mechanical treatment of alchemical changes. 
To this end, we mainly rely on the use of density functional theory (DFT) even though analogous arguments
can be made using conventional wave-function based quantum chemistry methods. 
In Sec.~\ref{sec:currentWork} we present numerical results that demonstrate the use
of alchemical derivatives for the screening of entire potential energy binding surfaces 
with semi-quantitative accuracy without additional self-consistent field calculations.

\section{Common alchemical applications}\label{sec:statistical}
Free energy is one of the most important state functions in chemistry. 
Since it is a statistical average, large numbers of configurations need to be taken into account to yield accurate predictions.~\cite{tuckerman_book_SM} 
E.g.,~calculating a free energy of solvation following a path that mimics the realistic complex
process of reversible microscopic immersion of the solute into a condensed ensemble of a very large number of solvent molecules
would imply a severe simulation effort that ensures that all relevant degrees of freedom have sufficiently been sampled. 
Furthermore, to account for hysteresis effects, this simulation should be repeated for various initial conditions and immersion rates. 
And one would have to start anew for any changes made to temperature, pressure, or solvent and solute species.
Alternatively, one could also calculate the {\em change} in free energy with respect to some solute
for which the free energy of solvation is already known. 
Thermodynamic integration, i.e.~numerical integration of the statistical mechanical average of the ``alchemical force'' 
along the path converting known solute ($\lambda = 0$) into query solute ($\lambda = 1$),\cite{Kollman_chemrev_1993}
\begin{eqnarray}
\Delta G = \int^1_0 d\lambda \left\langle \frac{\partial H(\lambda)}{\partial \lambda}\right\rangle_\lambda.
\label{eq:TP}
\end{eqnarray}

Jorgensen and Ravimohan\cite{Jorgensen_jcp_1985} proposed an even more efficient alternative:
One can also estimate the change in free energy of solvation due to changing the solute using perturbation theory and MC simulation. 
Specifically, they considered the effect on the free energy of hydration due to an alchemical change of a methyl into hydroxy-group, $\Delta G = G_f - G_i = G_{\mathrm{CH_3CH_3}} - G_{\mathrm{CH_3OH}}$.
One can show that if the sampling of the two states, $H_i$ and $H_f$, yields sufficient overlap, the corresponding
free energy difference can be accurately predicted using perturbation theory,
\begin{equation}\label{eq:TP}
	e^{-\beta \Delta G}  \approx \left\langle e^{\beta  (H_f - H_i)} \right \rangle_i.
\end{equation}
Here, $1/\beta = k_BT$, and the right-hand-side refers to the average of the Hamiltonian difference Boltzmann's weight over a trajectory generated using $H_i$. 
The authors used a linear interpolation of force field parameters for methanol and ethane,
$H(\lambda) = H_{\mathrm{CH_3OH}} - \lambda(H_{\mathrm{CH_3CH_3}} - H_{\mathrm{CH_3OH}})$,
from which the energy can be calculated for any $\lambda$. 

As such, alchemical changes enable the prediction of changes in free energy differences without having to actually model the realistic process under investigation. 
Linear interpolation approaches have been applied to free energy calculations in various chemical and biological systems.\cite{Jorgensen_science_2004, Gunsteren_pnas_2005, Oostenbrink_jpcb_2011, Leung_jctc_2011} 
Smith and van Gunsteren found that non-linear alchemical coupling not necessarily leads to linear free energy changes.\cite{Gunsteren_jpc_1994}
Further applications of alchemical coupling to the estimation of free energy difference include the free energy of hydration of ions using
{\em ab initio} molecular dynamics,\cite{AnatoleLeung_jcp_2009} 
differences in free energy of binding between various host-guest complexes,\cite{OostenbrinkHostGuest2008}
free energy differences at phase boundaries to predict melting points,\cite{Maginn_jpc_2007,sai-iecr2010}
the free energy of mixing to identify eutectics in ternary mixtures of molten alkali-nitrate salts,\cite{Anatole_pre_2011} 
kinetic isotope effects,\cite{Anatole_jctc_2011} 
as well as constraints on the composition of the Earth's core.\cite{Alfe_nature_2000}

But also from the solid state point of view the concept of alchemical coupling is used
for the prediction of properties of disordered materials, such as co-crystals, solid solutions, or solid mixtures, 
as a function of mole-fraction.\cite{disorderedMaterial} 
It is computationally difficult to deal with such mixed disordered systems since the minimal self-repeating units can become very large. 
As a result it is nearly impossible to set up disordered systems within periodic boundary conditions. 
One alternative consists of using cluster-expansion methods~\cite{ClusterCeder2009}, 
another alternative, akin to alchemical coupling, is the virtual crystal approximation (VCA)\cite{VCA} which 
averages the system, rather than explicitly representing the full system. 
One of the simplest disordered class of materials are ternary semiconductors, A$_x$B$_{1-x}$C, 
where AC and BC are two different semiconductors while $x$ is the mole-fraction between A and C. 
Consider, for example,\cite{Pettifor_2003} Eq.~(\ref{eq:H_lambda}) applied to Al$_x$Ga$_{1-x}$As: $H(x) = H_{\mathrm{GaAs}} + x (H_{\mathrm{AlAs}} - H_{\mathrm{GaAs}})$.
The linear interpolated alchemical path describes an averaged Hamiltonian between AlAs and GaAs for any mole-fraction of Al and Ga.

\section{Alchemy in Quantum Mechanics}\label{sec:QMAlchemy}
\subsection{Fictitious systems}
Within a first principles notion of CCS,\cite{Anatole_ijqc_2013} 
one can view every compound in any geometry as a state described by a unique Hamiltonian $H$. 
More specifically, the total potential energy's molecular Hamiltonian, $H$, is a function of a given set of 
nuclear coordinates, charges, and number of electrons, $\lbrace \mathbf{R}_I, Z_I, N_e\rbrace$, respectively. 
Without any loss of generality, we here rely on the Born-Oppenheimer approximation, neglecting all non-adiabatic electronic or nuclear quantum effects. 
Studies of alchemical paths have historically provided essential insight into the density functional theory (DFT) formulation of
the many-electron problem in molecules.~\cite{HK,parryang} 
In 1974, Harris and Jones introduced an adiabatic connection,\cite{Harris_adiabatic_1974} coupling the system of interest to an 
fictitious but relevant system of non-interacting electrons, 
\begin{equation}\label{eq:nonInteractE}
H(\lambda) = T + \lambda V_{ee} + V_{ext},
\end{equation}
where $T$, $V_{ee}$, and $V_{ext}$ represent kinetic energy, electron-electron interaction energy, and external potential energy operator. 
By changing $\lambda$ from 1 to 0, one can dial in the electron-electron interaction. 
For $\lambda = 0$, the electronic Schr\"odinger equation can thus be solved analytically, providing useful information on properties such as the exchange-correlation hole,\cite{Gunnarsson_prb_1976, Parr_DFT, Koch_DFT} 
an important ingredient for current exchange-correlation potential development efforts.\cite{Perdew_prb_1981, PBE, Levy_pra_1991, Perdew_prb_1992}
Another important study of electron-electron interaction, carried out by Seidl, Perdew and Levy, introduces the limit of strictly correlated electrons.\cite{Perdew_pra_1999} Replacing the variable $\lambda = \frac{1}{\mu}$ for $0<\mu\leq 1$ in Eq.~(\ref{eq:nonInteractE}), one obtains a coupled system where electron-electron interaction is dominant.

E.~B.~Wilson introduced the idea to alchemically couple any system to the uniform electron gas. 
Based on this path, he derived an expression for an exact four-dimensional density functional theory, integrating over three spatial and one $\lambda$-dimension.\cite{Parr_DFT, Wilson_jpc_1962} 
Subsequently, Politzer and Parr\cite{Parr_jpc_1974} showed that, by defining free-atom screening functions, Wilson's functional can be decomposed into kinetic and potential energy of $N_e$ electrons. These definitions of DFT related alchemical paths constitute the underlying framework for the results and discussions here within. 

Within DFT,\cite{HK} we can explicitly calculate $E(\lambda)$ 
for {\em any} iso-electronic change of geometry and composition, 
i.e.~under the constraint that $\int d\fatr \; n_\lambda(\fatr) = N_e \;\forall \;0 \le \lambda \le 1$,
\bea\label{eq:alchemyRCD}
		E[n_\lambda,\lambda] &=& T[n_\lambda] + V_{ee}[n_\lambda] + \int d\mathbf{r}\: n_\lambda(\mathbf{r})\:v_{ext}(\mathbf{r},\lambda).
\eea
Here, the coupling is introduced explicitly through the external potential.
In practice, such coupling can be realized by scaling up or down the pseudopotentials or nuclear charges
of initial and final molecules at their distinct clamped geometries. 
Note that kinetic and potential electron energy terms are only implicitly dependent on $\lambda$, 
namely through the electron density's dependency on the $\lambda$-dependent external potential---which is imposed through
application of the variational principle.

\subsection{Alchemical teleportation of an atom}
To illustrate the idea of alchemical changes within quantum chemistry, we now consider a process which is trivial
when done through a realistic path, and non-trivial when done alchemically: 
The ``teleportation'' of an atom from one site to another with the constraint that the total number of electrons and protons is kept constant. 
Thus, instead of the trivial real space displacement of the atom, we continuously decrease the nuclear charge (annihilation) 
at one site while continuously increasing (creation) the nuclear charge at the other site by the same amount. 
For example, the external potentials of an atom at two sites can be linearly coupled through an alchemical path,
\begin{equation}\label{eq:He2He}
 H(\lambda)= T + V_{ee} + Z \sum_i^{N_e}\Big(\frac{(1-\lambda)}{|\mathbf{r}_i|} +  \frac{\lambda}{|\mathbf{r}_i - \mathbf{R}|}\Big),
\end{equation}
where the respective atomic sites are located at the origin and at $\mathbf{R}$. 
Considering only the endpoints ($\lambda$ = (0,1)), the location of the atom obviously shifted from origin to $\mathbf{R}$. 
For any intermediate value of $\lambda$, however, the electrons will distribute among the two competing poles of the external potential
given in Eq.~(\ref{eq:He2He}), forming an attractive chemical bond.

To numerically exemplify this process, we have chosen the highest occupied molecular orbital (HOMO) eigenvalue, $\varepsilon$, as 
property of interest, and an alchemical change corresponding to the linear teleportation of a $Z=2$ and $N_e$ = 2 system, 
i.e.~effectively translating the He atom. 
The numerical calculation of $\varepsilon$ for variable $\lambda$ has been carried out 
using pseudopotential interpolation within plane-wave basis set PBE DFT calculations,
in analogy to previous studies.\cite{Anatole_jcp_2009, Anatole_prl_2005, Anatole_jcp_2010}
See Sec.~\ref{sec:compute} for more details. 

\begin{figure}
\includegraphics[scale=0.4, angle=0, width=8.5cm]{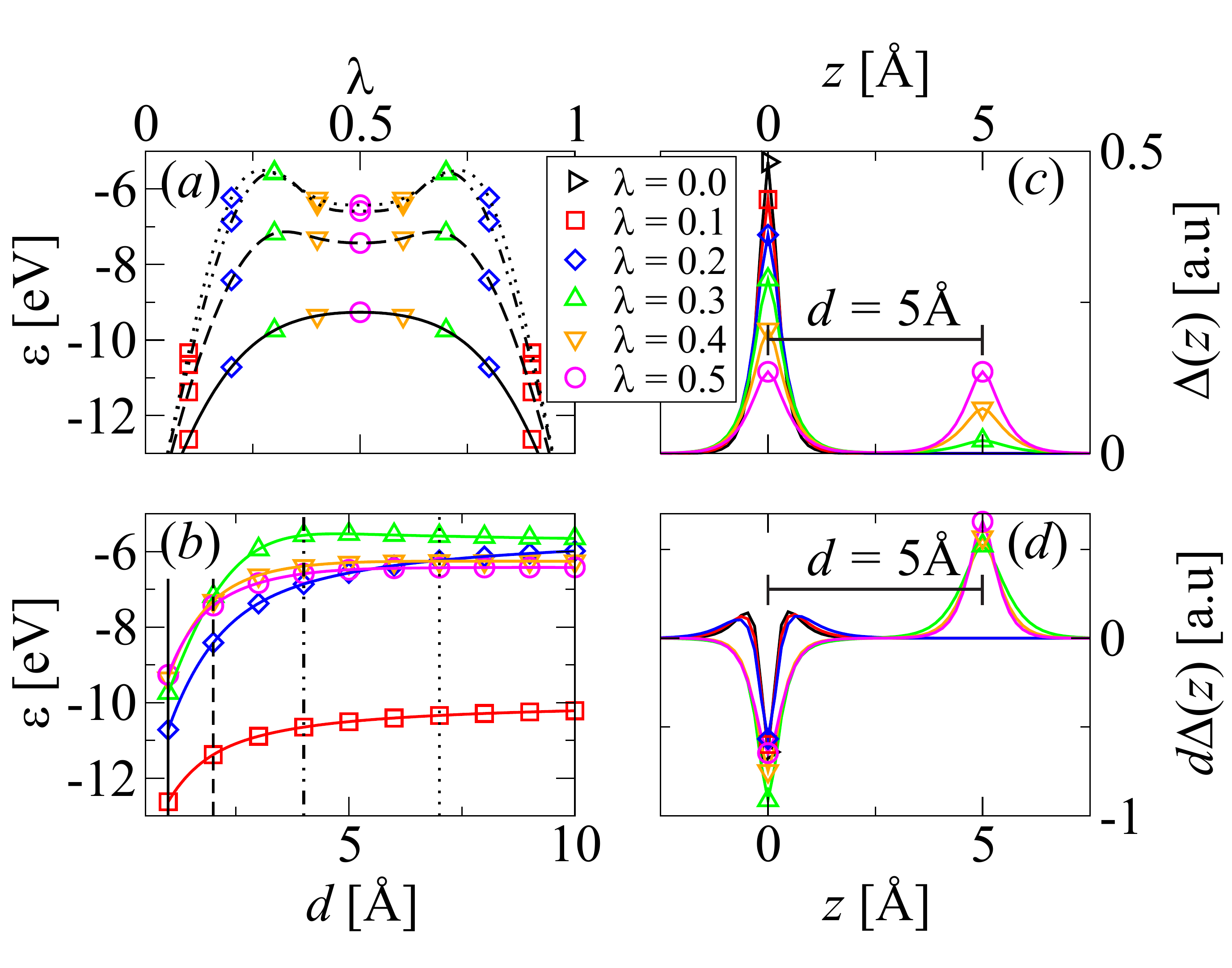}
\caption{
Alchemical transportation of a He atom.  (a) $\varepsilon$ as a function of $\lambda$ for various distances $d \in \{1,2,4,7\}${\AA} denoted by solid, dashed, dash-dotted, dotted, respectively.  
(b) $\varepsilon$ as functions of $d$.
(c) Integrated electron density, $\Delta(z)$ = $\int dxdy\;n_\lambda(x,y,z)$ for various $\lambda$ at $d = 5${\AA}. 
The electronic cusps at the nuclear sites have been highlighted by their corresponding $\lambda$ symbols. 
(d) Integrated response of electron density due to changing $\lambda$, $d \Delta(z)$ = $\int dxdy\:\partial_\lambda n_\lambda(x,y,z)$
for various $\lambda$ at $d = 5${\AA}.
}
\label{fig:He2He}
\end{figure}

In Fig.~\ref{fig:He2He}(a), the $\lambda$-dependence of $\varepsilon$ is shown for various distances between the two atomic sites, $d=|\mathbf{R}|$
Clearly, while alchemical paths for small $d$ yield simple parabolic shapes of $\varepsilon$, for teleportation involving 
larger interatomic distances $\varepsilon$ develops into a double hill. 
$\varepsilon$ versus $d$ is plotted in Fig.~\ref{fig:He2He}(b) for various $\lambda$ values. 
We note that for $\lambda$ = 0.5 (magenta), the $d$ dependency of $\varepsilon$ corresponds to the case of stretching H$_2$.
$\varepsilon$ increases monotonically at $\lambda=0.1$ and $\lambda=0.2$ as $d$ increases. 
For these $\lambda$ values, the buildup of integrated electron density at the $\mathbf{R}$, is still negligible, Fig.~\ref{fig:He2He}(c). 
Overall, the effect of nuclear potential in Eq.~(\ref{eq:He2He}), $\frac{2\lambda}{|\mathbf{r}_i - \mathbf{R}|}$, 
amounts to a static electric field, which induces static Stark effect.\cite{Griffiths_QM,StarkEffect_jcp_1999,StarkEffect_prb_2003} 
Because the electric field decreases according to Coulomb's law $\propto\frac{1}{d}$, 
$\varepsilon$ rises as a result of decreasing electric field perturbation.
Apart from the delocalization error of DFT,\cite{WYang_science_2008, WYang_cr_2012} 
such nonlinear behavior could also be related to the instability of H$_2^+$-like systems, 
which has been shown analytically.\cite{Baber_pcps_1935} 
Hogreve pointed out that strongly polarized electron density of asymmetric H$_2^+$-like molecule severely destabilizes the system.\cite{Hogreve_jcp_1993} 
While the additional electron stabilizes the system, nonlinear behavior can be expected for $\varepsilon$ in the case of strongly polarized density, i.e.~for $\lambda > 0.3$ (in Fig.~\ref{fig:He2He}(c)).
Fig.~\ref{fig:He2He}(c) displays integrated electron density slices, $\Delta(z) = \int dxdy\; n(x,y,z)$, 
for various $\lambda$ values at interatomic distance, $d = 5${\AA}.
Note that for $\lambda= 0.5$, the electron density distribution corresponds to H$_2$. 
The non-linear dependency of electron density $n$ on linearly changing growth of nuclear charge can be seen 
in Fig.~\ref{fig:He2He}(d)
for the abrupt changes in electron density response induced by going from $\lambda \approx 0.2$ to $\lambda \approx 0.3$.
To investigate the impact of parameterized exchange correlation potentials in DFT,
Cohen and Mori-S\'anchez calculated similar changes for $N_e =1$ and $N_e = 2$ using the hydrogen atom 
plus one additional atomic site where a nuclear charge is grown, 
i.e.~$Z(\lambda)$ with $Z(\lambda = 0) = 0$, $Z(\lambda = 0.5) = 1$ (H), and $Z(\lambda = 1) = 2$ (He).~\cite{CohenDramaticChanges2014}

\section{Rational Compound Design}\label{sec:currentWork}
\subsection{Motivation}
The goal of rational compound design (RCD) corresponds to solving the inverse question, 
i.e.~``which compounds exhibit a set of pre-defined desired properties?'', at a rate that is superior
to mere screening.\cite{Beratan_1996, Beratan_science_1991, Zunger_1999, Anatole_prl_2005, Reiher_ijqc_2014} 
Various approaches tackle this problem, including the inverse spectrum approach,\cite{Reiher_njc_2007} 
linear combination of atomic potentials,\cite{WYang_jacs_2006, WYang_jcp_2007} and many others.\cite{Zunger_1999, Wener_prl_1994, Zunger_prb_1995, Jansen_acie_2002} 
For the electronic potential energy,
an alchemical path coupling $E(\lambda = 0)$ of one molecule to an unknown $E(\lambda=1)$ 
of another compound makes explicit the compositional dependence of the energy. 
Understanding such a dependence holds promise to dramatically reduce the computational 
burden of having to stubbornly screen one compound after the other. 
More specifically, we can expand $E$ in $\lambda$ in terms of a Taylor series, 
\begin{equation}\label{eq:Taylor}
	E(\lambda) = E_i + \lambda \partial_{\lambda} E_i + \frac{\lambda}{2}\partial_{\lambda}^2 E_i + \cdots,
\end{equation}
where the subscript of $E_i$ represents the quantum mechanical expectation value of $H_i$. 
In other words, if all derivatives of $E_i$ were available one could simply follow
a steepest descent procedure to screen a set of coupled ``neighboring'' molecules, 
{\it e.g.}~with small differences in geometry or stoichiometry,
to identify and proceed to more promising compound candidates.
Fig.~\ref{fig:alchemicalPath}(b) illustrates the exploration of CCS following such alchemical predictions. 
Ideally, only a single calculation of the electronic ground-state $E_i$ would be required (denoted by black circle). 
The energy of neighboring compounds can then be estimated via Eq.~(\ref{eq:HFDerivative}) (denoted by white circles). 
As we discuss below, it is possible to make such scans through changes in geometry as well as composition.

In Ref.~\cite{Anatole_jcp_2009} we already discussed that for {\em any} iso-electronic alchemical change, 
the first order derivative is simply the Hellmann-Feynman derivative.~\cite{Feynman_pr_1939}
Consequently, differentiation of Eq.~(\ref{eq:alchemyRCD}) yields,
\bea
\label{eq:HFDerivative}
	\partial_{\lambda} E[n_\lambda,\lambda] &=& \left\langle\partial_{\lambda} H\right\rangle_\lambda = \int d\mathbf{r}\:n_\lambda(\mathbf{r})\:\partial_{\lambda} v(\mathbf{r},\lambda),
\eea
which is the same as the first order perturbation term.\cite{Griffiths_QM} 
Higher order derivatives can be evaluated or approximated by linear response theory,\cite{WeitaoY_pra_1988, DFPT,Sebastiani_2000} 
and will be discussed below in the context of linearizing the energy in $\lambda$ in Sec.~(\ref{sec:Linear}).


\subsection{Alchemical changes in geometry}
We now consider alchemical changes that only involve teleportation. 
To demonstrate the versatility and transferability of the discussed approach,
we have calculated alchemical predictions of changes in binding energy for two very
different modes of binding: The covalent interatomic potential in hydrogen fluoride, 
as well as the hydrogen-bond-dominated van der Waals potential of the water dimer.
In both cases the binding energy is given as the difference in potential energy
of dimer (dim) and (relaxed) monomers m1 and m2, $E_b(d) = E_{\mathrm{dim}}(d) - E_\mathrm{m1} - E_\mathrm{m2}$.
Any approximate solution of the electronic Schr\"odinger equation at some initial distance $d_i$ enables
us to estimate the binding energy of any other $d$ using the Hellmann-Feynman derivative 
and first order Taylor expansion in the alchemical 
teleportation path (Eqs.~(\ref{eq:HFDerivative},\ref{eq:Taylor})), 
\begin{equation}\label{eq:T1}
E_b(d) \;\; \approx \;\; E_b^{T1}(d) \;\; = \;\; E_b(d_i) + \partial_{\lambda} E_b(d_i).
\end{equation}

\begin{figure}
\centering
\includegraphics[scale=0.4, angle=0, width=8.5cm]{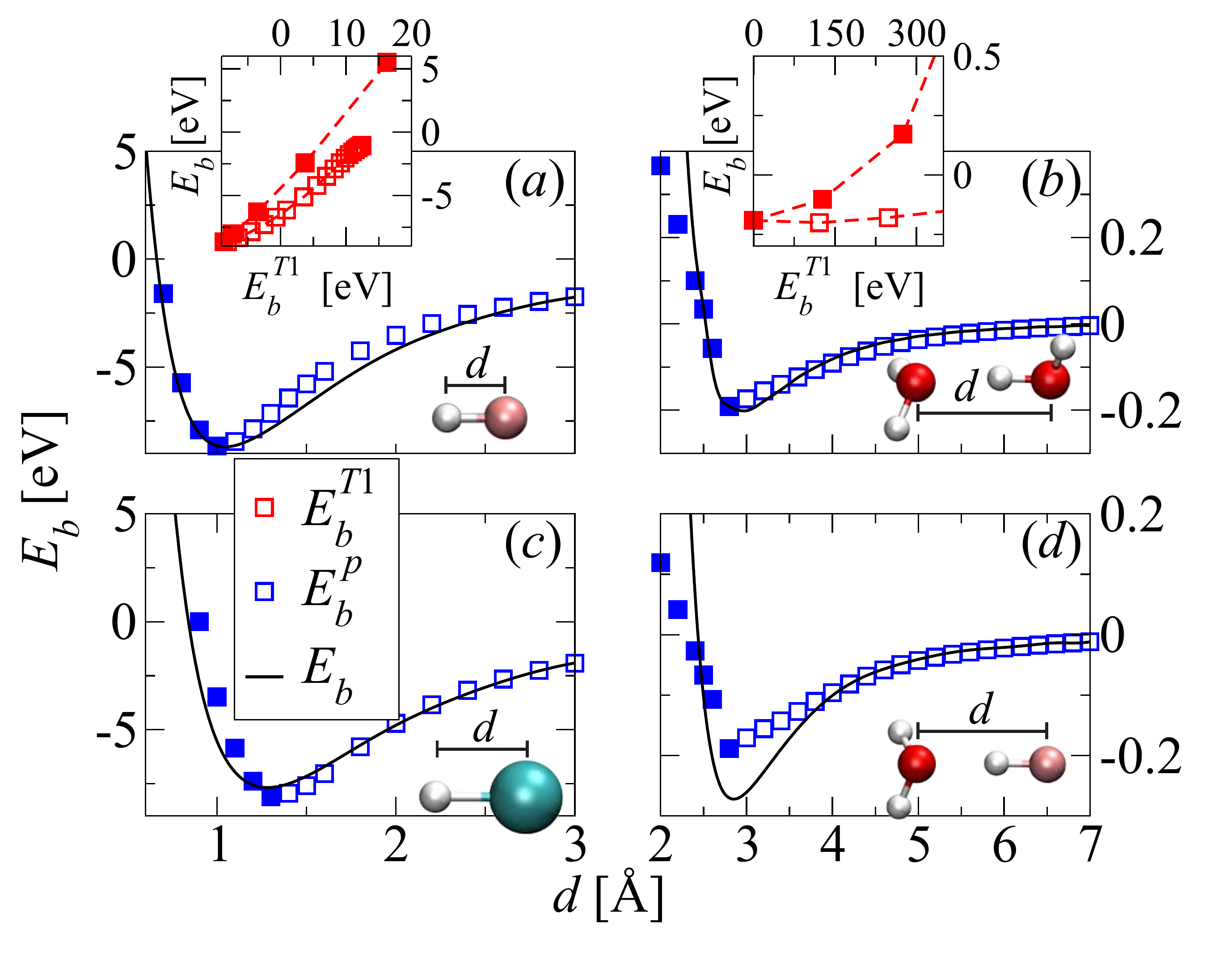}
\caption{
Actual (black lines) and alchemical (blue squares) binding energy $E_b$ of repulsive (filled) and attractive (empty) regions of binding potentials for
HF (a), (H$_2$O)$_2$ (b), HCl (c), and H$_2$O-HF (d). 
Each screen corresponds to using only one self-consistent field (SCF) calculation at $d_i = d_{eq}$, together
with the first order Taylor-expansion based model, $E^p_b = a_{l/r} E_b^{T1} + b_{l/r}$ (Eq.~(\ref{eq:Taylor})).
Insets in (a) and (b) show $E_b$ versus $E_b^{T1}$. 
The screens in (c) and (d) are slightly less predictive because they are made using
SCF results from HF and (H$_2$O)$_2$, respectively.
$d_{eq}$ is set to 1, 2.8, 1.4 and 2.8 \AA\ for (a), (b), (c), and (d) respectively.
}
\label{fig:Curves}
\end{figure}

Considering now the case of $d_i$ corresponding to the equilibrium distance, $d_{eq}$,
the insets of the two top panels in Fig.~\ref{fig:Curves} show the resulting
scatter plots of $E_b^{T1}(d)$ versus the actual $E_b(d)$ for various values of $d$ in the
case of HF and (H$_2$O)$_2$.
While there is clear correlation, the scale differs dramatically for the two modes of binding. 
Most importantly, in the case of the dissociative tail $E^{T1}$ correlates practically linearly 
with the actual binding energy.
Consequently, if we now approximate the true $E_b \approx E_b^p = a_{l/r} E_b^{T1} + b_{l/r}$, 
($l$ and $r$ correspond to the left-hand repulsive wall and the right-hand attractive tail, 
respectively)
one can solve for the coefficients if further constraints are known.
Since this is a rather exploratory study, we here simply assume that
(i) $E_b(d=d_{eq}) = E_b^{T1}(d_{eq})$,
and (ii) $E_b(d\rightarrow\infty) = 0$ in the dissociative region of the curve,
and (iii) in the case of the repulsive region that $E_b(d=\frac{2}{3}d_{eq}) = 0$ for covalent binding, 
and $E_b(d=\frac{5}{6}d_{eq}) = 0$ for intermolecular binding.
Assumption (iii) is based on experience using typical Morse and Lennard-Jones parameters.
All resulting coefficients $\{a_{l/r},b_{l/r}\}$ are specified in Ref.~\cite{ParameterResults}.
The predictions for scanning the entire binding potential agree reasonably well with the true binding potentials,
and are shown together for both systems in the  top panels in Fig.~\ref{fig:Curves}.
Integrated deviations of these predictions are also shown in Table~\ref{tab:MAE},
yielding single digit percentage error for predicting the integral over the covalent bonding 
potential of hydrogen fluoride, and $\sim$14\% error for the integral over the van der Waals 
potential of the water dimer.
We stress that the {\em entire} screen using this model only requires a single self-consistent field
cycle to calculate energy and derivatives at $d = d_{eq}$. 

\begin{table}
\caption{
Numerical integrals of reference energies $E_b$ (REF) and of absolute deviation of 
alchemical predictions $E_b^p$ from reference energies $E_b$ (PRE-REF) over 
the binding region, i.e. for all $d$ where $E_b < 0$, 
and percentage thereof (\%)
for the repulsive wall predictions as well as for the attractive tail.
Columns correspond to 
(a) HF, (b) (H$_2$O)$_2$, (c) HCl, and (d) H$_2$O-HF on display in Fig.~(\ref{fig:Curves}),
}
\label{tab:MAE}
\begin{tabular}{|c|c|c|c|c|} \hline
Integral [eV$\times$\AA]           &  (a)   & (b)  & (c)  & (d) \\ \hline
REF [eV$\times$\AA] (wall)     & -2.542 & -0.059 & -3.054& -0.095 \\
PRE-REF [eV$\times$\AA] (wall) &  0.150 &  0.009 & 0.656 & 0.036 \\ 
\% (wall) &  5.9    & 15.6    & 21.5   & 37.6 \\ \hline
REF [eV$\times$\AA] (tail)     & -8.199 & -0.275 & -6.017& -0.594 \\
PRE-REF [eV$\times$\AA] (tail) &  0.692 & 0.031  & 0.239 & 0.072 \\ 
\% (tail) &  8.4    & 11.4    & 4.0    & 12.1 \\ \hline
\end{tabular}
\end{table}

\subsection{Alchemical changes in stoichiometry}
We now extend the use of Eq.~(\ref{eq:T1}) to also make predictions not only for teleportation changes in geometry but also for
transmutational changes in stoichiometry. 
In particular, we have calculated predictions for changing hydrogen fluoride into hydrogen chloride at various distances,
as well as changing the water dimer into the water-hydrogen fluoride complex.
Since we use pseudopotentials for both of these changes the total number of valence-electrons in our calculations does not change. 
To calculate $E^{T1}_b$ according to Eq.~(\ref{eq:T1}) we have chosen $d_{eq}$ to correspond to the equilibrium distance
of the target system, i.e.~HCl and H$_2$O-HF. 
Again, the same assumptions (i)-(iii) as above are used to calculate $a_{l/r}$ and $b_{l/r}$ to
obtain a linear approximation of the actual $E_b(d)$ in $E^{T1}_b$.
Also for these changes, the resulting coefficients are specified in Ref.~\cite{ParameterResults}.
The predicted binding curves show reasonable agreement with the actual numbers, as shown for both 
systems in the bottom panels in Fig.~\ref{fig:Curves}.
Again, integrated and relative errors are given in Table~\ref{tab:MAE}, and show a reasonable
albeit slightly worse performance than in the case of predicting the water dimer or the hydrogen fluoride.
We reiterate, however, that the entire screen results from only one self-consistent field cycle carried out to 
calculate energy and derivative of {\em another} molecular system---at the $d_{eq}$ of the target system. 
While it is also possible to use other $d$ to calculate energies and derivatives this typically
leads to less accurate predictions. 
We do not think that this constitutes a problem since knowledge about equilibrium distances of 
target structures can easily be obtained from literature or through
inexpensive force-field or semi-empirical quantum chemistry calculations 
which incur negligible computational overhead.

\subsection{Linearizing chemical space}
\label{sec:Linear}
As we have seen above for the teleportation of the He atom, as well as in other studies,\cite{Anatole_jcp_2009, Anatole_ijqc_2013} 
there are cases when the first order Taylor expansion of Eq.~(\ref{eq:T1}) does not provide satisfactory predictive power. 
This is not surprising since changes in composition correspond to large perturbations that typically lead to non-linear responses. 
We believe that the good performance obtained above for the binding curves is due to cancellation of
higher order effects and due to the calibration of the linear model to the appropriate physical dissociation or repulsion limits.
One way to systematically improve the predictive accuracy consists of including increasingly higher-order terms. 
Sebastiani and coworkers\cite{Sebastiani_2000, Sebastiani_2012, Sebastiani_2013} as well as 
Geerlings, De Proft and others\cite{Geerlings_jctc_2000, Geerlings_jcpa_2013, Geerlings_csr_2014} 
proposed promising approaches in this direction. 
For example, akin to our discussion above, Benoit, Sebastiani and Parrinello investigated the performance of second order linear response 
theory for screening the potential energy surface of the water dimer, and achieved very high predictive power.~\cite{dbdsmp}
How to efficiently calculate susceptibility accurately and in general, however, is still a matter of current research. 
Furthermore, typically one observes a (sometimes dramatic) increase in computational cost due 
to wave function-dependent susceptibilities, 
thereby defying the original motivation of RCD to navigate CCS {\em without} having to solve 
Schr\"odinger's equation from scratch for each and every new geometry or molecule.
As pointed out in Ref.~\cite{Anatole_jcp_2009}, a promising alternative route towards improving the 
predictive power of the first order derivative consists of deviating from the assumption that the
alchemical coupling must be linear in $\lambda$.
In fact, as already mentioned above in the context of interpolating force-fields,~\cite{Gunsteren_jpc_1994}
we are free to use {\em any} kind of coupling as long as we meet our endpoints, 
i.e. comply with Eq.~(\ref{eq:H_lambda}).
More specifically, if we knew the form of some coupling external potential $v_{ext}(\fatr,\lambda)$ 
that induces such changes in the electron density that $E(\lambda)$ becomes linear in $\lambda$, 
then Eq.~(\ref{eq:T1}) would result in perfect predictions.
The quest for such a potential has been discussed in Ref.~\cite{Anatole_ijqc_2013}, 
in particular in connection to a 1-ounce-of-gold prize for anyone who provides a solution to this problem. 

For a coupling path to generally fulfill the requirement that $E(\lambda)$ becomes linear in $\lambda$ we note that 
the potential must have such a shape that the first order derivative, 
$\partial_\lambda E$ is a constant (as already pointed out and used in Ref.~\cite{Anatole_jcp_2009}),
and that furthermore, all higher order energy derivatives must be zero.
Consequently,
\bea
0 & = & \partial^m_\lambda E \;\; = \;\; \int d\fatr \; \partial^{m-1}_\lambda (n_\lambda(\fatr) \partial_\lambda v_{ext}(\fatr,\lambda)),
\label{eq:Constraint}
\eea
$\forall \; m > 1$.
This imposes certain constraints on the interpolating potential. 
For example, in the case of the second order derivative, 
equating the integrand to zero and solving for the electron density's response results in
\bea
\partial_\lambda n(\fatr) & = & -n_\lambda(\fatr)\frac{\partial^2_\lambda v_{ext}(\fatr,\lambda)}{\partial_\lambda v_{ext}(\fatr,\lambda)}.
\eea
Similar expressions can be obtained for higher order density response functions. 
Possibly, Eq.~(\ref{eq:Constraint}) could be transformed into a variational problem
that yields an interpolating potential with the desired effect that the
associated energies that are indeed linear in $\lambda$.


\section{Conclusions}
We discussed recent theoretical developments and approaches based on coupling states using unrealistic ``alchemical'' paths. 
Numerical evidence has been presented for the applicability and versatility of alchemical approaches applied to the inexpensive 
prediction of quantum mechanical observables of novel systems. 
The derivative based predictions certainly reflect the qualitative trend of the
desired binding potentials, and are accurate within single, or low double, digit percentage accuracy. 
Results, discussions, and current state of the field indicate that the study of 
generalized coupling approaches still holds great promise for the predictive simulation of
molecular and materials properties, as well as for rational compound design. 

\section{Acknowledgments}
Both authors acknowledge funding from the Swiss National Science foundation (No.~PPOOP2\_ 138932).
This research used resources of the Argonne Leadership Computing Facility at Argonne National Laboratory,
which is supported by the Office of Science of the U.S.~DOE under contract DE-AC02-06CH11357.

\section{Computational Details}\label{sec:compute}
All calculations have been carried out using Kohn-Sham DFT\cite{KSDFT} as implemented in {\tt CPMD}\cite{CPMD} 
with PBE (He teletransportation) or PBE0 (all other calculations) functional\cite{PBE, PBE0}. 
Goedecker pseudopotentials\cite{Goedecker_1996, Goedecker_1998, Goedecker_2013} have been used 
as published by Krack,\cite{Krack_2005}, in conjunction with
100 Ry plane-wave cutoffs in isolated $30\times 15\times 15\:$\AA$^3$ box for He, 
$20\times 20\times 20\:$\AA$^3$ for HF$\rightarrow$HF and HF$\rightarrow$HCl, 
and 110 Ry plane-wave cutoff with isolated $25\times 15\times 15\:$\AA$^3$ box for H$_2$O$\rightarrow$H$_2$O and H$_2$O$\rightarrow$HF.
Alchemical coupling has been imposed through linear interpolation of corresponding pseudopotential parameters\cite{Anatole_jcp_2009, Anatole_prl_2005, Anatole_jcp_2010} $\sigma(\lambda) = \sigma_i + \lambda (\sigma_f - \sigma_i)$, where $\sigma_i$ and $\sigma_f$ represent the parameters for atoms in $H_i$ and $H_f$ respectively. 
HOMO eigenvalues $\varepsilon$ have been calculated as a finite difference relying on Jana`k and Koopman's theorem,\cite{Koopman_1987, WeitaoY_prb_2008} $\varepsilon \approx \frac{E_{N+\delta} - E_N}{\delta}$,
where $\delta$ is 1\% of a positive unit charge. 
The geometry scans of HF and HCl have been performed by fixing the heavy atoms at origin, moving H in $d$ direction, while aligning the HF or HCl bond along $d$-axis. In the case of the (H$_2$O)$_2$ scan, all geometries have been relaxed, setting the oxygen of the H-acceptor at the origin, while aligning the O-H bond of the H-donor with the $d$-axis. 
The H$_2$O-HF geometry scans have been performed by replacing the oxygen of the H-donor by F and annihilating the other hydrogen while keeping the HF bond aligned with the $d$-axis. 

\bibliography{literature}{}
\bibliographystyle{ieeetr}

\end{document}